\documentclass[seceq]{ptptex}
\usepackage{graphics,epsf,bm}
\def\bK{\mbox{\boldmath $K$}}
\def\bR{\mbox{\boldmath $R$}}
\def\bk{\mbox{\boldmath $k$}}

\def\br{\mbox{\boldmath $r$}}
\def\bs{\mbox{\boldmath $s$}}
\def\ssc{$S_{\mbox{\footnotesize SCDW}}(W)$ }
\def\sgf{$S_{\mbox{\footnotesize GF}}(W)$ }

\title{$\Xi$-nucleus potential and $(K^-,K^+)$ inclusive spectrum at $\Xi^-$
production threshold region}

\author{Michio \textsc{Kohno}$^1$ and Shintaro \textsc{Hashimoto}$^2$
}

\inst{$^1$Physics Division, Kyushu Dental College,
Kitakyushu 803-8580, Japan\\
$^2$Advanced Science Research Center, Japan Atomic Energy Agency,
Ibaraki 319-1195, Japan
}

\abst{The $(K^-,K^+)$ $\Xi^-$ production inclusive spectrum is reinvestigated
in view of the very weak $\Xi$-nucleus potential predicted by microscopic
calculations with the SU$_6$ quark-model baryon-baryon interaction.
The inclusive spectrum is evaluated by the semiclassical distorted wave
(SCDW) method. The explicit comparison of the strength function with
that of the Green-function method demonstrates the quantitative
reliability of the SCDW approximation. It is presumed that the presently
available data at the $\Xi$ production threshold region does not
necessarily imply the attractive strength of about 15 MeV
for the $\Xi$-nucleus potential in a conventional Woods-Saxon form.
Instead, an almost zero potential is preferable.
}

\begin{document}

\maketitle

\section{Introduction}
The study of the baryon-baryon interactions has made steady progress toward the
$S=-2$ sector. The construction of a $\Lambda N$ interaction model is almost
under control, based on the experimental data of $\Lambda$ hyper nuclei.
The $\Sigma N$ interaction was elusive until 1990s because there was
no clear $\Sigma$ bound states observed in $\Sigma$ formation spectra, except
for the $^4_\Sigma$He due to the attraction in the specific isospin $T=1/2$
channel. Now it seems to be established that the $\Sigma$-nucleus potential
is repulsive with the weak attraction at the nuclear surface. Such property
was suggested first by the analyses of the $\Sigma^-$ atomic level shifts by
Batty, Friedman, and Gal \cite{BFG94}. The overall repulsive nature of the
$\Sigma$-nucleus potential was indicated by the $(\pi^-,K^+)$ $\Sigma^-$
inclusive spectra at KEK \cite{KEK}. This feature of the $\Sigma$-nucleus
interaction is predicted by the microscopic calculations \cite{MK00}
starting from the SU$_6$ quark-model interactions \cite{QMBB}. Then,
the interactions in the $S=-2$ sector, either $\Lambda\Lambda$ or $\Xi N$,
is the forefront of the experimental studies of the strangeness nuclear
physics. This knowledge is naturally invaluable for the quantitative
predictions for various aspects of neutron star matter.

There have been a few experimental data for the properties of the $\Xi$
hyperon in the nuclear medium. The old emulsion data was used by Dover
and Gal \cite{DG83} to conclude that the depth of the $\Xi$-nucleus
potential was more than 20 MeV. At present, there are only a few sets
of experimental cross sections for the $\Xi^-$ production by $(K^-,K^+)$
reactions. One is the experiment at KEK by Iijima \textit{et al.} \cite{II92},
that covers a wide range of the outgoing $K^+$ energy. The other is the data by
Fukuda \textit{et al.} \cite{TF98} and Khaustov \textit{et al.} \cite{KH00},
in which the chief motivation is the search of the $\Xi$ bound state
and thus the measurements were concentrated on the
energy region around the $\Xi$ production threshold. Both data sets are
insufficient to establish the $\Xi$-nucleus interaction, because of the
energy and angle resolutions as well as the limited number of counts.
Nevertheless, these data around the $\Xi$ production threshold were
analyzed to indicate that the depth is 16 MeV and 14 MeV, respectively.
This magnitude seems to be canonical at present for the $\Xi$-nucleus
potential. The estimation relies solely on anlyses by means of distorted wave
impulse approximation (DWIA) reported in Refs. 7) and 8).
Those theoretical calculations of the $\Xi$ formation inclusive spectra, however,
contain various simplifying approximations and uncertainties, which hinders
the reliability of discussing absolute value of the cross sections.
It is necessary to reanalyze the data by an independent calculational framework.

The microscopic calculations in Ref. 9) with the SU$_6$ quark-model
baryon-baryon interactions fss2 \cite{QMBB} predict that the localized
$\Xi$-nucleus potential in finite nuclei  fluctuates around 0 inside a nucleus
with some weak attraction at the nuclear surface region. Observing that
the $\Sigma$-nucleus potential calculated microscopically in Ref. 9)
shows a good correspondence with the empirical character without adjustments,
the prediction for the $\Xi$ is also credible. It is interesting to investigate
whether such a weak $\Xi$-nucleus potential provides $\Xi$ formation spectra
consistent with the experimental data.

In this paper, we examine the data of Ref. 8) by employing the SCDW
method \cite{LK90} for evaluating the spectrum in which the energy and angle
dependences of the elementary cross section are respected \cite{KF06}. The
nuclear Fermi motion is properly taken into account by the Wigner
transformation of the target s.p. wave functions. The distorted waves of
the incoming and outgoing kaons are described by the Klein-Gordon equation.
Actually we reported results of the SCDW calculation for $(K^-,K^+)$
inclusive spectra in Ref. 12). In that paper we mainly referred to the
data by Iijima \textit{et al.} \cite{II92}. Here, the attention is focused
on the data around the $\Xi$ production threshold.

In Sec. II, we compare the SCDW method and the Green-function method
frequently used for evaluating the inclusive spectrum, taking the
$(K^-,K^+)$ $\Xi^-$ production process as an explicit example. The
comparison shows that the SCDW method is reliable to discuss
quantitatively the cross section. Calculated results are compared with
the data by Khaustov \textit{et al.} \cite{KH00} in Sec. III.
A summary is given in Sec. IV.
 
\section{Comparison of the SCDW method with the Green-function method}
A Green-function method has been widely used for analyzing various hadron
production inclusive spectra. The use of the Green function for treating
final state interactions in inclusive reactions was presented, for example,
in Ref.13) for the inclusive $(e,e')$
reactions. The application to hyperon production processes was initiated by
the study of the production of $\Sigma$-hypernuclear states in $(K^-,\pi^+)$
reaction by Morimatsu and Yazaki \cite{MY88}. The calculation for inferring
the $\Xi$-nucleus potential depth to be about $14$ MeV on the basis of the
$(K^-,K^+)$ events on carbon \cite{KH00} essentially employs the same method,
although the strength function is actually calculated by the Kaper-Peierls
method \cite{MOT}.

In this section we compare our SCDW method with the Green-function method
in numerical detail. The basic formula of the double differential cross
section for the inclusive $(K^-,K^+)$ $\Xi^-$ production reaction
in a distorted wave impulse approximation is
\begin{equation}
 \frac{d^2 \sigma}{dW d\Omega} =\frac{\omega_{i,red}\omega_{f,red}}{(2\pi )^2}
 \frac{p_f}{p_i} \sum_{p,h}\; \frac{1}{4\omega_i \omega_f} | \langle \chi_f^{(-)} \phi_p^{(-)} |v_{f,p,i,h}|
 \chi_i^{(+)} \phi_h \rangle |^2
 \delta (W-\epsilon_p + \epsilon_h ),
\end{equation}
where $\chi_i^{(+)}$ and $\chi_f^{(-)}$ represent the incident $K^-$ and
final $K^+$ wave functions with energies $\omega_i$ and $\omega_f$,
respectively, and $W=\omega_i -\omega_f$ is the energy transfer. The
corresponding momenta are represented by $p_i$ and $p_f$.
The reduced energy with respect to the target nucleus (residual hyper
nucleus) is denoted by $\omega_{i,red}$ ($\omega_{f,red}$).
The formula describes the process in which the nucleon in the occupied
single-particle state $h$ is converted to the unobserved outgoing $\Xi$
hyperon state $p$. The elementary amplitude
of the process $K^- + p \rightarrow K^+ +\Xi^-$ is denoted
by $v_{f,p,i,h}$, which depends on the energy and momentum of the
particles in the reaction.

The summation $\sum_p$ is taken over the complete set of the unobserved
final $\Xi$ hyperon states. This summation with the energy-conserving
$\delta$-function can be written by the Green function $G(\br,\br';W)$.
\begin{eqnarray}
 \sum_p \delta (W-\epsilon_p + \epsilon_h )
 \langle \br|\phi_p^{(-)} \rangle \langle\phi_p^{(-)}|\br'\rangle
 = -\frac{1}{\pi} Im \sum_p
 \frac{\langle \br|\phi_p^{(-)}\rangle\langle\phi_p^{(-)}
 |\br'\rangle}{W-\epsilon_p +\epsilon_h+i\epsilon}
 \nonumber \\
 = -\frac{1}{\pi} Im \sum_p
 \frac{\langle \br|\phi_p^{(+)}\rangle\langle\phi_p^{(+)}
 |\br'\rangle}{W-\epsilon_p +\epsilon_h+i\epsilon}
 = -\frac{1}{\pi} Im\; G(\br,\br';W).
\end{eqnarray}
Note that this expression can be extended to the case that the $\Xi$ hyperon
is described by a complex optical model potential to take care of the
decaying processes to other channels. In our SCDW method, $\phi_p^{(-)}$
is described by a real potential. Effects of the inelastic channels are
taken into account by convoluting the calculated spectrum by a Lorentz-type
distribution function. 

The Green-function method commonly introduces a factorization approximation;
otherwise practical calculations are hard to be carried out. Namely, the
elementary amplitude $v_{f,p,i,h}$ is replaced by some average and taken out
of the integration and the summation. In this case, the differential cross
section becomes
\begin{equation}
 \frac{d^2 \sigma}{dW d\Omega} =\beta \left( \frac{d \sigma}{d\Omega}\right)_{av}
 S_{\mbox{\footnotesize GF}}(W),
\end{equation}
introducing the factor $\beta$ which represents the difference of the kinematics
in the two-body and $A$-body systems. In actual calculations, a recoil correction
is incorporated. Denoting $\zeta\equiv \frac{A-1}{A}$ with $A$ being the mass
number of the target nucleus, the strength function \sgf is defined by
\begin{equation}
 S_{\mbox{\footnotesize GF}}(W)=-\frac{1}{\pi} Im
 \langle \chi_f^{(-)*}(\zeta\br) \chi_i^{(+)}(\zeta\br) \phi_h(\br)|
 G(\br,\br';W)|\chi_f^{(-)*}(\zeta\br') \chi_i^{(+)}(\zeta\br') \phi_h(\br')\rangle.
\end{equation}
The determination of $\beta$ and $\left( \frac{d \sigma}{d\Omega}\right)_{av}$
admits ambiguities.

The SCDW model does not introduce the factorization approximation. Instead,
the wave function between two points $\br$ and $\br'$ is approximated by
a plane wave with the local classical momentum $\bk(\bR)$:
\begin{equation}
 \chi_{i,f}^{(+)}(\br) \chi_{i,f}^{(+)*}(\br') 
 = \chi_{i,f}^{(+)}\left(\bR+\frac{1}{2}\bs\right) \chi_{i,f}^{(+)*}
 \left(\bR-\frac{1}{2}\bs\right)
 \simeq |\chi_{i,f}^{(+)}(\bR)|^2 e^{i\bk(\bR)\cdot\bs}.
\end{equation}
The direction of $\bk(\bR)$ is calculated by the quantum mechanical
momentum density $\bk_q(\bR)$
\begin{equation}
 \bk_q(\bR)= \frac{\mbox{Re} \{\chi^{(\pm)*}(\bR)(-i)\nabla\chi^{(\pm)}(\bR)\}}
 {|\chi^{(\pm)}(\bR)|^2},
\end{equation}
and the magnitude of $\bk(\bR)$ is determined by the energy-momentum
relation at $\bR$
\begin{equation}
 m_K^2 + \bk^2(\bR) +2 \omega_{i,f} (U_R(\bR)+V_{Coul}(\bR))
 -V_{Coul}^2(\bR)=\omega_{i,f}^2, 
\end{equation}
where $m_K$ is the kaon mass, $V_{Coul}$ is the Coulomb potential, and
$U_R(\bR)$ is the real part of the
optical potential for $\chi_{i,f}$ with the energy $\omega_{i,f}$. The
expression of the double differential cross section in this approximation is
explained in Ref. 11) and the inclusion of recoil corrections is
specified in Ref. 12). The final expression reads
\begin{eqnarray}
\frac{d^2 \sigma}{dW d\Omega} = \frac{\omega_i \omega_f}{(2\pi )^2}
 \frac{p_f}{p_i} \xi^6 \int \int d\bR d\bK \sum_{p}\;
 \frac{1}{4\omega_i \omega_f} |\chi_f^{(-)}(\bR)|^2
 |\chi_i^{(+)}(\bR)|^2 \nonumber \\
 \times |\phi_p^{(-)}(\xi \bR)|^2  |v_{f,p,i,h}|^2 \frac{(2\pi)^3}{\xi^3}
 \sum_h \Phi_h\left(\xi\bR,\frac{1}{\xi}\bK\right) \nonumber \\
 \times \delta(\bK +\bk_i(\bR)-\bk_f(\bR) -\bk_p(\bR))\delta (W-\epsilon_p + \epsilon_h ),
\end{eqnarray}
where $\xi=\frac{1}{\zeta}= \frac{A}{A-1}$ appears by taking care of the
recoil effects and $\Phi_h$ is the Wigner transformation of the density matrix
of the nucleon hole state wave function. If we introduce the factorization
approximation, we can define the strength function \ssc in the SCDW method.
\begin{eqnarray}
 S_{\mbox{\footnotesize SCDW}}(W)= \xi^6 \int \int d\bR d\bK \sum_{p}\;
 |\chi_f^{(-)}(\bR)|^2 |\chi_i^{(+)}(\bR)|^2 |\phi_p^{(-)}(\xi \bR)|^2 \frac{(2\pi)^3}{\xi^3}
 \nonumber \\
  \times  \sum_h \Phi_h\left(\xi\bR,\frac{1}{\xi}\bK\right)
 \delta(\bK +\bk_i(\bR)-\bk_f(\bR) -\bk_p(\bR))\delta (W-\epsilon_p + \epsilon_h ).
\end{eqnarray}

\begin{figure}[b]
 \begin{center}
 \epsfxsize=0.6\textwidth
 \epsffile{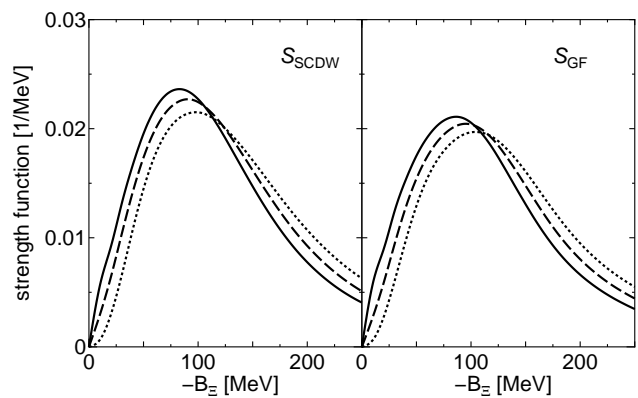}
 \caption{Strength functions in the case of the $(K^-,K^+)$ $\Xi^-$
production inclusive reaction on $^{12}$C at $p_{K^-}=1.8$ GeV/c.
The solid, dashed, and dotted curves correspond to the three choices of
the strength of the $\Xi$-nucleus potential in a standard Woods-Saxon form
($r_0=1.2\times 12^{1/3}$ fm  and $a=0.65$ fm);  $U_\Xi^0=-20,\;-5,\;+10$ MeV,
respectively. The left panel shows the results of the SCDW method,
\ssc, and the right panel those of the GF method, \sgf.
}
\end{center}
\end{figure}

While the energy and angle dependences of the elementary amplitude $|v_{f,p,i,h}|^2$
are, in the SCDW method, treated explicitly inside the integration of Eq. $(2\cdot 8)$,
though the on-shell approximation has to be used, the approximation of Eq. $(2\cdot 5)$
brings about certain uncertainties. The SCDW approximation has been successful
in quantitatively describing intermediate energy $(p,p'x)$ and $(p,p'n)$ inclusive
reactions \cite{OG}. However, we cannot expect
a priori that the replacement of Eq. $(2\cdot 5)$ is always reliable. As noted above,
if we set $|v_{f,p,i,h}|^2=1$, we obtain the strength function in the SCDW treatment,
which can be directly compared with the exact strength function \sgf in the
Green-function method. It is useful to assess the reliability of the SCDW
method by comparing \ssc with \sgf by numerical calculations.

We show, in Fig. 1, the strength functions \ssc and \sgf for
the $(K^-,K^+)$ $\Xi^-$ production at $p_{K^-}=1.8$ GeV/c which are obtained
from the three choices of the strength of the $\Xi$-nucleus potential,
$U_\Xi^0=-20,\;-5$, and $+10$ MeV, in a standard Woods-Saxon
form ($r_0=1.2\times A^{1/3}$ fm and $a=0.65$ fm). The density-dependent
Hartree-Fock wave functions of Campi and Sprung \cite{CS} are used for the
hole states of $^{12}$C. The $K^-$ and $K^+$
distorted waves are provided by solving the Klein-Gordon equation with the
optical potential parameters given in Ref. 12). There are bound states
in $s$ and $p$ orbits for the case of  $U_\Xi^0=-20$ MeV, but those contributions
are not shown in Fig. 1. The \ssc agrees well with the
\sgf at all energies. The difference is seen at most 10 \%. On the basis of this
correspondence, we are assured to use the SCDW method to include the energy
and angle dependences of the elementary process in the nuclear medium together
with the explicit treatment of the nucleon Fermi motion.

Note that if we multiply the strength function by
$\beta \left(\frac{d\sigma}{d\Omega}\right)_{av}$, we readily obtain double
differential cross sections in a factorization approximation, although the energy
dependence of the multiplicative factor is not simple to determine
when discussing the yield over the wide range of excitation energies.

\section{SCDW-model calculations of $(K^-,K^+)$ $\Xi^-$ production inclusive
spectrum on carbon}

\begin{figure}[t!]
 \begin{center}
 \epsfxsize=0.6\textwidth
 \epsffile{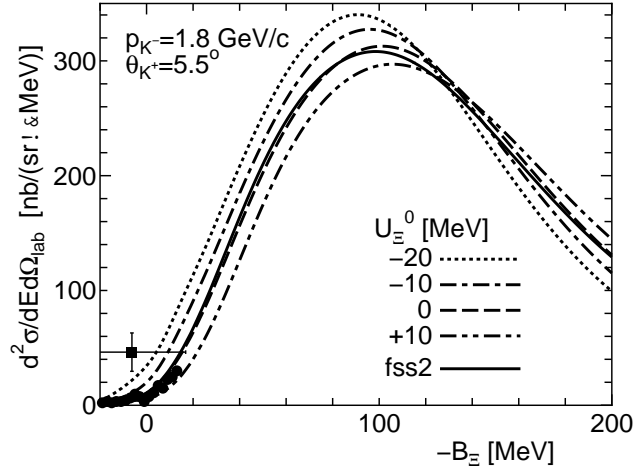}
\end{center}
 \caption{$(K^-,K^+)$ $\Xi^-$ production inclusive spectra on $^{12}$C at
$p_{K^-}=1.8$ GeV/c. The solid  curve is the calculated result at
$\theta_{K^+}=5.5^\circ$ for the $\Xi$-nucleus potential suggested by the
quark-model $\Xi N$ interaction fss2. Other curves are the results at
$\theta_{K^+}=5.5^\circ$ for potentials in a single Woods-Saxon form with the
depths of $+10$, $0$, $-10$, and $-20$ MeV, respectively, assuming standard
geometry parameters of $r_0= 1.2\times 12^{1/3}$ fm and $a=0.65$ fm.
The data by Khaustov \textit{et al.} \cite{KH00} is shown by filled circles.
Note that the experimental data is the average of the cross sections over
the angles between $0^\circ$ and $8^\circ$ and the original cross section
in Ref. 8) is given as a histogram with a 2 MeV step. The data by
Iijima \textit{et al.} \cite{II92} at the threshold region is also shown
by a filled square, though the incident $K^-$ momentum is 1.65 GeV/c.}
\end{figure}

\begin{figure}[t!]
 \begin{center}
 \epsfxsize=0.6\textwidth
 \epsffile{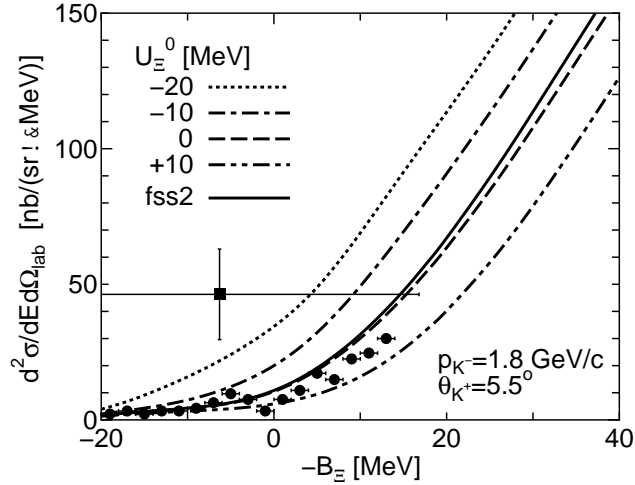}
\end{center}
 \caption{Magnification of the threshold region of Fig. 2.
}
\end{figure}

We show, in Fig. 2, the calculated spectra by the SCDW model
for $(K^-,K^+)$ $\Xi^-$ production inclusive reactions on $^{12}$C
at $p_K=1.8$ GeV/c, corresponding to the experiment by
Khaustov \textit{et al.} \cite{KH00} at KEK. The threshold region
is magnified in Fig. 3. Because the experimental data is the average
cross section between the angles of $0^\circ$ and $8^\circ$,
the calculations carried out at $\theta_{K^+}=5.5^\circ$ are shown.
This angle is the mean value in the following meaning. The angle average
of the cross section $\sigma(\theta)$ over the angle between $\theta_{min}$
and $\theta_{max}$ is given by
\begin{equation}
 \sigma^{av}(\theta_{min},\theta_{max})
 =\frac{\int_{\theta_{min}}^{\theta_{max}} \sigma(\theta ) \sin\theta d\theta}
  {\int_{\theta_{min}}^{\theta_{max}}  \sin\theta d\theta} .
\end{equation}
Because the scattering angle is limited to the forward region, it is
sufficient to adopt the approximation $\sin\theta \sim \theta$ and
$\sigma(\theta) \sim \sigma(0)+\sigma'(0)\theta +\frac{1}{2}
\sigma''(0)\theta^2$. The mean angle $\theta_{av}$ that gives
$\sigma(\theta_{av}) =\sigma^{av}(0^\circ, 8^\circ)$ is in the vicinity
of $5.5^\circ$, irrespective of $\sigma(0)$, $\sigma'(0)$, and $\sigma''(0)$.

In the SCDW method, the wave function of the unobserved $\Xi$ hyperon
$\phi_p^{(-)}$ is described by a real optical model potential. The effects
of inelastic channels are incorporated by convoluting the calculated spectrum
with a Lorentz-type distribution function. We assign 2 MeV to the half width
$\Gamma /2$, which is consistent with the empirical indication by the $\Xi^- p$
elastic and inelastic cross section measurements at low energy by
Ahn \textit{et al.} \cite{AHN06} and also the weak imaginary part of the
$\Xi$-nucleus potential obtained from the microscopic calculation with
the quark-model potential \cite{KF09}. In addition, we take into account
the experimental resolution of $\Delta E= 6.1$ MeV \cite{KH00}
by smearing the spectrum with a Gaussian function:
\begin{equation}
 f(E,\Delta E) = \frac{1}{\Delta E} \sqrt{\frac{\log 2}{\pi}}
 e^{-\log 2\left( E/\Delta E \right)^2}.
\end{equation}

Calculations are carried out for the $\Xi$-$^{12}$C potential parametrized in
a sum of two Woods-Saxon forms on the basis of the microscopic calculations
in Ref. 9). The parameters are tabulated in Table I. We
also evaluate, for the purpose of reference, the spectrum with potentials
described by a single Woods-Saxon form in a standard geometry of
$r_0=1.2\times 12^{1/3}$ fm and $a=0.65$ fm with
$U_\Xi^0=+10$, $0$, $-10$, and $-20$ MeV, respectively.
The $K^- +p\rightarrow K^+ +\Xi^-$ elementary differential cross section is
taken from the parametrization by Nara \textit{et al.} \cite{NA97}. We show,
in Fig. 4, the angle dependence of the elementary cross section of this
parametrization in the laboratory frame for two incident $p_{K^-}$ momenta
of 1.8 GeV/c and 1.65 GeV/c. Note that although the differential cross
section in the center-of-mass frame is backward peaking, in the laboratory
frame the cross section in the forward angles is larger. 

\begin{table}
\caption{Strength and geometry parameters of the Woods-Saxon form
$f_i(r)=U_i^0 /[1+\exp((r-r_{0,i})/a_i)]$ fitted to the real part of the
$\Xi$ single-particle in $^{12}$C calculated microscopically \cite{KF09}
with the quark-model potential fss2.}
\begin{center}
\begin{tabular}{cccc} \hline\hline
 $i$ & \hspace*{0.5em}$U_i^0$ (MeV)\hspace*{0.5em}
      & \hspace*{0.5em}$r_{0,i}$ (fm)\hspace*{0.5em}
      & \hspace*{0.5em}$a_i$ (fm) \\ \hline
  1 & $-4.139$ & 3.569 & 0.5291 \\
  2 & $10.53\;$  & 2.130 & 0.3032 \\ \hline \hline
\end{tabular}
\end{center}
\end{table} 

\begin{figure}[t]
 \begin{center}
 \epsfxsize=0.6\textwidth
 \epsffile{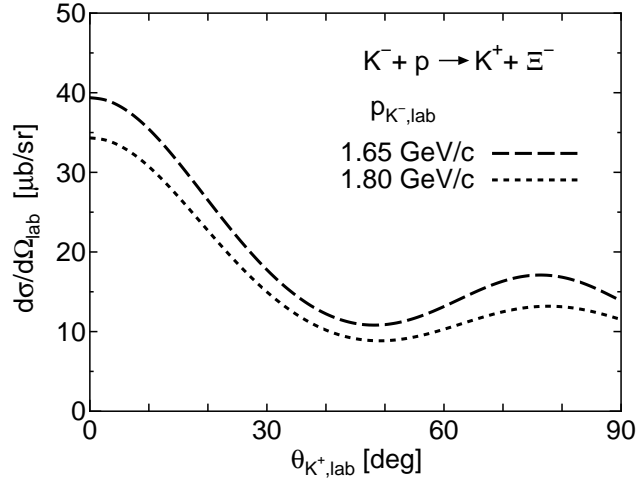}
 \caption{Angle dependence of the $K^- +p\rightarrow K^+ +\Xi^-$
elementary cross section parametrized by Nara \textit{et al.} \cite{NA97}
in the laboratory frame at $p_{K^-}=1.65$ GeV/c and $p_{K^-}=1.8$ GeV/c,
respectively.
}
\end{center}
\end{figure}

The potential based on the quark-model potential fss2 gives a very similar
result to the case of $U_\Xi^0=0$. Calculations with different strengths
of the $\Xi$-nucleus potential show that the peak position and the width
of the spectrum change systematically. An attractive potential,
even a weak one as $U_\Xi^0=-10$ MeV, is seen to predict larger cross
sections at the $\Xi$ production threshold region than the experimental
data. Thus, an almost zero potential is preferable to account for
data by Khaustov \textit{et al.} \cite{KH00}.

The same data was analyzed in Ref. 8) to judge that the depth of
the $\Xi$-nucleus potential is 14 MeV or less. Because the details of the
theoretical calculation are not found in Ref. 8), it is hard to
figure out the cause of the difference from our result. In Fig. 5, we
compare the spectrum before folding the experimental resolution by the
SCDW model calculation with that given by Eq. $(2\cdot 3)$ of the Green-function
method for the two cases of the $\Xi$-nucleus potential in a Woods-Saxon
form. In this case, the strength function \sgf is evaluated with
including the imaginary potential. The absorptive strength in the
same Woods-Saxon form as the real part is set to be 4 MeV. Considering
surface effects, this strength is regarded to correspond
to $\frac{\Gamma}{2}=2$ MeV for smearing the spectrum of the SCDW model.
In Ref.7) for the $(K^-,K^+)$ reaction at $p_{K^-}=1.65$ GeV/c,
$\beta \left(\frac{d\sigma}{d\Omega}\right)_{av}$ in Eq. (2.3) is set
as $0.73\times (35\pm5)\sim 26$ $\mu$b/sr. The average differential cross
section at forward angles of $35\pm5$ $\mu$b/sr is consistent with the cross
section shown in Fig. 4. However, if we use $\beta=0.73$, the spectrum
overestimates the experimental data. To obtain the similar strength as the
SCDW cross sections around the $\Xi$ production threshold, we need
$\beta \left(\frac{d\sigma}{d\Omega}\right)_{av}= 0.33\times 35$ $\mu$b/sr.
It is seen in Fig. 5 that if the multiplicative average cross section is
assumed to be energy independent, the spectrum of the Green-function
method tends to underestimate the cross section at around the peak position.

\begin{figure}[t!]
 \begin{center}
 \epsfxsize=0.6\textwidth
 \epsffile{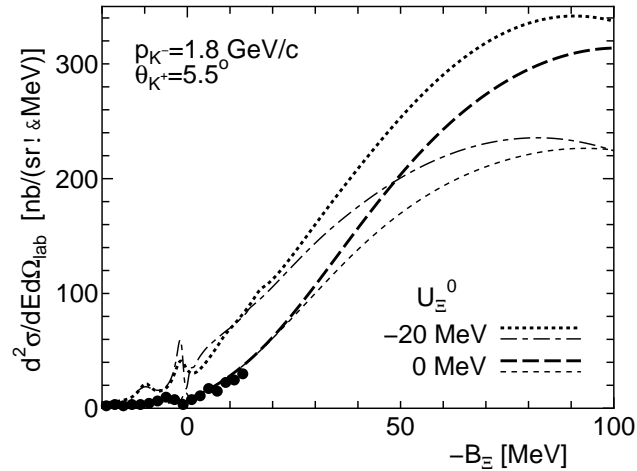}
\end{center}
 \caption{$(K^-,K^+)$ $\Xi^-$ production inclusive spectrum on $^{12}$C
at $p_{K^-}=1.8$ GeV/c calculated by the SCDW method (thick curves) is
compared with that of the Green-function method (thin curves) without including
the experimental resolution. The latter spectrum is given by Eq. $(2\cdot 3)$
with $\beta\left(\frac{d\sigma}{d\Omega}\right)_{av}=0.33\times35$ $\mu$b.
\sgf is evaluated with including the absorptive potential of the strength
of 4 MeV. The number of $\beta=0.33$ is chosen to match with the cross
sections of the SCDW method at the threshold region. Note that $\beta=0.73$
is used in Ref. 8) for $p_{K^-}=1.65$ GeV/c.}
\end{figure}

As for the other $(K^-,K^+)$ $\Xi^-$ production data by
Iijima \textit{et al.} \cite{II92}, the SCDW-model calculation accounts only
the half of the experimental magnitude, as reported in Ref. 12).
Other theoretical calculations so far reported show similar underestimation.
The result on carbon target from the intra-nuclear cascade model
calculation by Nara \textit{et al.} \cite{NA97} is similar to our spectrum.
The DWIA calculation with the Green-function method by Tadokoro \textit{et al.}
\cite{TAD} seemingly agrees well with the experimental data. However,
they present only the spectrum at $\theta_{K^+}=0^\circ$. Although they
commented that the angle dependence was negligibly small, this is questionable
because our calculations both in the SCDW method and the Green-function
method as well as the calculation in Ref. 8) indicate strong angle
dependence of the $(K^-,K^+)$ $\Xi^-$ production spectrum. If we consider the
spectrum at $\theta_{K^+}=0^\circ$, our SCDW result is close to that of
Tadokoro \textit{et al.}

The change of the strength of the $\Xi$-nucleus potential shifts the peak
position and alters the width of the spectrum, but it cannot increase the
height of the spectrum by a factor of 2. If the magnitude of the elementary
amplitude is reliable, we presume that there is inconsistency between the
data at $p_{K^-}=1.65$ GeV/c by Iijima \textit{et al.} \cite{II92} which
covers a wide range of the $\Xi$ excitation energy and the data at
$p_{K^-}=1.8$ GeV/c by Khaustov \textit{et al.} \cite{KH00} which is
limited to the energy region around the $\Xi^-$ production threshold.

\section{Summary}
We have examined whether the weak $\Xi$-nucleus potential suggested
by microscopic calculations \cite{KF09} using the SU$_6$ quark-model
baryon-baryon interaction fss2 \cite{QMBB} provides a reasonable description
for $\Xi$ formation spectra available at present. We employ the SCDW model
to calculate the $(K^-,K^+)$ $\Xi^-$ production spectrum. This method
was developed \cite{KF06} for evaluating various inclusive cross sections,
respecting the energy and angle dependences of the elementary process
together with the nuclear Fermi motion. The semiclassical local momentum
approximation is demonstrated, in this paper, to be sufficiently
reliable to discuss quantitatively cross sections by explicitly comparing
the SCDW strength function and the precise strength function calculated
in the Green-function method.

In the present calculation we use the $\Xi$-nucleus potential as
parameterized in a sum of two Woods-Saxon forms on the basis of the
microscopic calculation \cite{KF09} in $^{12}$C. We do not claim that the
potential is definitely accurate and reliable, but regard it as the typical
weak potential with non-monotonic radial dependence which does not
support any $\Xi$ nuclear bound state. The microscopic calculation of
the $\Xi$-nucleus potential with the quark-model $\Xi N$ interaction
fss2 indicates that the imaginary strength is rather small,
which is consistent with the experimental data \cite{AHN06}. We employ
the half width of 2 MeV to smear the SCDW spectrum using the Lorentz-type
distribution function. To compare the calculated result with the data,
we fold, in addition, the experimental resolution of 6.1 MeV \cite{KH00}
by a Gauss functional form. We also calculate the spectrum with the
$\Xi$-nucleus potentials, $U_\Xi^0=+10,\;0,\;-10$, and $-20$ MeV,
in a standard Woods-Saxon form to examine the potential
dependence of the $\Xi$ production cross section.

Our conclusion is that the weak $\Xi$-nucleus potential yields a right
order of magnitude of the experimental $\Xi$ production cross section
available at present. If we use a standard Woods-Saxon form, the strength
of the $\Xi$-nucleus potential should be almost zero. This result does not
agree with the speculation by the DWIA analyses in Ref. 8) that
the well depth is about $14$ MeV. We have to bear in mind, however, that this
experimental data is based on the small number of counts that are a few
tens or fewer and thus may not be accurate enough to conclude the strength
of the $\Xi$-nucleus potential. More experimental data with a better
resolution and statistics will be obtained from the $(K^-,K^+)$ experiment
being prepared in the J-PARC project \cite{JPARC}. Our
calculations show that even if the experiment finds little $\Xi$ production
strength in the $\Xi$ bound state region, we are able to infer the strength
of the $\Xi$-nucleus potential. This will promote our understanding of the
baryon-baryon interactions in the $S=-2$ sector and improve theoretical
models of them.


\end{document}